\documentclass[12pt,a4paper]{article}
\usepackage{jheppub}
\usepackage{dsfont}
\usepackage{amssymb,amsmath}
\usepackage{epsfig}
\usepackage{epstopdf}
\usepackage{latexsym}
\usepackage{graphicx}
\usepackage{subfigure}
\usepackage{booktabs}
\usepackage{bbm}
\pdfoutput=1
\makeatletter
\def\@fpheader{\relax}
\makeatother

\def\O{{\cal O} }
\def\C{{\cal C} }

\def\S{{\cal S} }
\def\L{{\cal L} }

\def\H{{\cal H} }
\def\M{{\cal M} }

\def\dcut{{%
    \setbox0\hbox{D}%
    \rlap{\hbox to \wd0{\hss ~/ \hss}}\box0}}

\subheader{\begin{flushright}
UTTG-25-12\\
TCC-024-12
\end{flushright}}

\title{Holographic Mutual Information at Finite Temperature}
\author{Willy Fischler,$^{a,b}$~Arnab Kundu,$^a$~Sandipan Kundu$^{a,b}$}
\affiliation{$^a$Theory Group, Department of Physics, University of Texas, Austin, TX 78712}
\affiliation{$^b$Texas Cosmology Center, University of Texas, Austin, TX 78712}
\emailAdd{fischler@physics.utexas.edu}
\emailAdd{arnab@physics.utexas.edu}
\emailAdd{sandyk@physics.utexas.edu}

\abstract{Using the Ryu-Takayanagi conjectured formula for entanglement entropy in the context of gauge-gravity duality, we investigate properties of mutual information between two disjoint rectangular sub-systems in finite temperature relativistic conformal field theories in $d$-spacetime dimensions and non-relativistic scale-invariant theories in some generic examples. In all these cases mutual information undergoes a transition beyond which it is identically zero. We study this transition in detail and find universal qualitative features for the above class of theories which has holographic dual descriptions. We also obtain analytical results for mutual information in the specific regime of the parameter space. This demonstrates that mutual information contains the quantum entanglement part of the entanglement entropy, which is otherwise dominated by the thermal entropy at large temperatures.}

\begin{document}

\maketitle
\flushbottom

\section{Introduction}

The gauge-gravity duality\cite{Banks:1996vh, Maldacena:1997re} is a concrete realization of the holographic principle\cite{Susskind:1994vu} which states that the number of degrees of freedom in quantum gravity scales like area. This idea, that was conceived from the nature of black hole entropy, {\it entangles} ideas in quantum gravity and information theory. In recent years, there has been a great progress in understanding aspects of strongly coupled large $N$ gauge theories using the AdS/CFT correspondence and generalizations thereof.

Within this context, {\it i.e.} analyzing systems described by such large $N$ gauge theories, it is an intriguing possibility to implement ideas that are natural in quantum information theory. One such important concept is entanglement entropy, which at zero temperature, measures the quantum entanglement between two sub-systems of a given system. In a quantum field theory, entanglement entropy of a region $A$ contains short-distance divergence which also scales like the area\cite{Bombelli:1986rw, Srednicki:1993im}. For large $N$ theories, which in the gravity dual are described by classical Einstein gravity with suitable matter fields, entanglement entropy can be computed using the Ryu-Takayanagi conjectured formula proposed in \cite{Ryu:2006bv, Ryu:2006ef}. This conjectured formula does indeed satisfy many non-trivial relations\cite{Headrick:2007km, Headrick:2010zt} known in quantum information theory. There have been numerous works analyzing entanglement entropy in various systems that are described by such classical gravity dual backgrounds, see {\it e.g.} \cite{Nishioka:2009un, Takayanagi:2012kg} for recent reviews.

However, due to its short distance divergence structure, entanglement entropy is a scheme-dependent quantity. This issue can be avoided by introducing an appropriate linear combination of entanglement entropies, which introduces a new concept named mutual information: $I(A,B) = S_A + S_B - S_{A\cup B}$, where $S_Y$ denotes the entanglement entropy of the region $Y$. Mutual information is an important concept in information theory that has certain advantages over entanglement entropy. It is (i) finite, (ii) positive semi-definite, (iii) measures the total correlations between the two sub-systems $A$ and $B$ and (iv) it is proportional to the entanglement entropy when $B \equiv A^c$, where $A^c$ denotes the complement of $A$, such that $S_{A\cup A^c} =0$.\footnote{We are assuming that $A\cup A^c$ is in a ground state with no degeneracy. Also, in order for $B$ to become $A^c$, we necessarily need $B$ to approach $A$. When two region $A$ and $B$ approach each other, new divergences appear depending on the shape of $A$ and $B$; see \cite{Swingle:2010jz} for more details on this. This is precisely the short-distance divergence structure observed in entanglement entropy.} Moreover, it can be proven\cite{PhysRevLett.100.070502} that mutual information satisfies an area law at finite temperature. This is to be contrasted with the behaviour of entanglement entropy which is dominated by the thermal entropy at large temperatures and hence follows a volume law.  Thus it is expected that mutual information carries more relevant content as far as describing quantum entanglement is concerned, since entanglement is still expected to scale as the are rather than the volume.

It was pointed out in \cite{Headrick:2010zt}, that in holographic duals, mutual information does undergo a ``first order phase transition" as the separation between the two rectangular sub-systems $A$ and $B$ is increased. For small separation, $I(A,B) \not= 0$, but for large separation $I(A,B) =0$; in the bulk there are always two candidate minimal area surfaces for the computation of $S_{A\cup B}$ and depending on the separation of $A$ and $B$ one or the other is favoured.\footnote{This is explained in more details in the next section.} Clearly, this does not correspond to a phase transition in the usual sense; however when $I(A,B)=0$, the two sub-systems $A$ and $B$ become completely decoupled. Hence we will call it a ``disentangling transition".

In this article, we study this disentangling transition for large $N$ relativistic conformal theories in $d$-spacetime dimensions in the presence of a finite temperature. The corresponding ``phase diagram" can be presented in the $(x/l)$ vs $(xT)$ plane, where $l$ denotes a linear size of the rectangular regions $A$ and $B$ (which we take to be of equal size for simplicity), $x$ is the separation between them and $T$ denotes the temperature of the system. Using the analytical methods developed in \cite{Fischler:2012ca}, we also explore the finite temperature behaviour of mutual information for this class of theories. We then move on to analyzing the same disentangling transition in non-relativistic scale invariant theories, {\it e.g.} for Lifshitz and hyperscaling-violating backgrounds. We find that this disentangling transition has universal qualitative features for all such theories with holographic duals.

It has been suggested in recent years that the emergence of a holographic space can be envisioned from the entanglement properties of a large class of many body quantum systems at criticality, see {\it e.g.} \cite{swingle:0905}. In the presence of a high temperature, entanglement entropy is dominated by thermal entropy and classical correlations. Mutual information, on the other hand, subtracts out the thermal contribution and still satisfies an area law. Furthermore, in an appropriate regime of parameters we analytically demonstrate that the finite piece of the mutual information actually captures the sub-leading term in the entanglement entropy at large temperature, and hence is a better guide to capturing quantum entanglement.

Perhaps one key universal feature alluded to in a couple of paragraphs earlier is the fact that mutual information decreases monotonically for increasing temperature. Hence, in this context, disentanglement in the boundary theory corresponds to raising the temperature, which in the dual gravitational picture can be viewed as the extremal surface probing deeper in the bulk. This is to be contrasted with the situation described in \cite{Czech:2012be}, where the emergence of an asymptotically globally AdS spacetime is described in terms of entangled states of a pair of CFTs defined on a hyberbolic space. It is nonetheless suggesting a possible deep connection between disentanglement and emergence of a holographic direction.

This article is divided in the following parts: in the next section we introduce the concept of entanglement entropy and mutual information more formally and discuss the holographic prescription. Section 3 is devoted to the discussion of properties of mutual information at finite temperature in various limits of $(x/l)$ and a numerical study of the disentangling transition for CFT$_d$. In section 4, we continue analyzing similar physics in non-relativistic scale-invariant theories by considering generic examples of Lifshitz and hyperscaling-violating backgrounds. Finally we conclude in section 5. Several details relevant for obtaining analytical results for mutual information at finite temperature have been relegated to three appendices.

\section{Entanglement entropy, mutual information and a summary}

In this section we will briefly elaborate on the definitions of entanglement entropy and mutual information. We will also include a brief summary of the results that we will discuss in the subsequent sections.

Let us begin with the ideas of entanglement entropy and mutual information. Consider a $d$ (spacetime) dimensional quantum field theory (QFT). Quantum systems are described by state vectors $| \psi \rangle \in \H$, where $\H$ denotes the Hilbert space of the system, evolving with some Hamiltonian $H$. A quantum system is also described by the density matrix, usually denoted by $\rho$, and defined as: $\rho =  | \psi \rangle \langle \psi |$. The expectation value of an operator $\O$ is then simply obtained by $\langle \O \rangle = {\rm tr} \left(\rho \O\right)$. Also, note that the entropy of such a system is given by the von Neumann formula: $S = - {\rm tr} \left[ \rho \log \rho \right] $.

Now let us consider a QFT defined on $\M^{d-1,1}$: a Lorentzian manifold.\footnote{For our current purposes we will focus only on Minkowski space: $\mathbb{R}^{d-1,1}$.} On a constant time Cauchy surface let us imagine dividing the system in two sub-systems, $A$ and $A^c$ respectively, where $A^c$ is the complement of $A$. The total Hilbert space then factorizes: $\H = \H_A \otimes \H_{A^c}$. We can define a ``reduced" density matrix of the sub-system $A$ by tracing out the information contained in $\H_{A^c}$ and thus define 
\begin{eqnarray}
\rho_A = {\rm tr}_{A^c} \left[ \rho \right] \ ,
\end{eqnarray}
and subsequently define the von Neumann entropy described by
\begin{eqnarray}
S_A = - {\rm tr} \left[ \rho_A \log \rho_A \right] 
\end{eqnarray}
as the entanglement entropy. Entanglement entropy is proportional to the number of degrees of freedom residing on the boundary shared by the sub-systems $A$ and $A^c$. The leading order divergence thus follows an area law\cite{Bombelli:1986rw, Srednicki:1993im}\footnote{We also note that the area law has violations in physically interesting and important cases. One simple example is $(1+1)$-dimensional conformal field theory, where a logarithmic violation arises. There is a simple scaling intuition behind such area law and its violations\cite{Swingle:2010jz}.}
\begin{eqnarray}
S_A = \alpha \frac{\partial A}{\epsilon^{d-2}} + \ldots \ ,
\end{eqnarray}
where $(\partial A)$ denotes the area of the region $A$ and $\epsilon$ denotes an UV cut-off of the QFT (in the limit of $\epsilon \to 0$); in a discretized version of the QFT, this cut-off can be identified with the lattice spacing. The constant $\alpha$ depends on the regularization scheme and thus is not universal.

It is a challenging task to compute entanglement entropy in a given quantum field theory. Within the realm of the AdS/CFT correspondence, more generally the holographic principle, there is a particularly simple yet powerful proposal for computing entanglement entropy for strongly coupled theories. The proposal was given in \cite{Ryu:2006bv, Ryu:2006ef} for static backgrounds and later generalized in \cite{Hubeny:2007xt} for backgrounds with explicit time-dependence. For a recent review, see {\it e.g.}~\cite{Nishioka:2009un, Takayanagi:2012kg}. According to this proposal, entanglement entropy of region $A$ is given by the Ryu-Takayanagi formula
\begin{eqnarray} \label{eedef}
S_A = \frac{{\rm Area} \left(\gamma_A\right)}{4 G_N^{(d+1)}} \ ,
\end{eqnarray}
where $G_N^{(d+1)}$ is the Newton's constant in $(d+1)$ bulk dimensions, $\gamma_A$ denotes the $(d-1)$-dimensional minimal\footnote{$\gamma_A$ is extremal in case the background has explicit time-dependence, as described in \cite{Hubeny:2007xt}.} area surface whose boundary coincides with the boundary of the region $A$: $\partial \gamma_A = \partial A$ and we also require that $\gamma_A$ is homologous to $A$. As described in \cite{Ryu:2006bv, Ryu:2006ef}, the Ryu-Takayanagi formula has passed several non-trivial checks.

At finite temperature, the corresponding ``reduced" density matrix can be defined as: $\rho_A = e^{-\beta H_A}$,\footnote{Note that, even at zero temperature one can write the reduced density matrix $\rho_{\rm reduced} = e^{- \hat{H}}$, where $\hat{H}$ is some hermitian operator referred to as the ``modular Hamiltonian" in \cite{Haag:axiom}.} where the total Hamiltonian of the system can at least be schematically represented as: $H = H_A + H_{A^c} + H_\partial $. Here $H_A$ and $H_{A^c}$ denotes the Hamiltonians of the sub-systems $A$ and $A^c$ respectively; $H_\partial$ denotes the interactions between the two sub-systems across the boundary.\footnote{Strictly speaking, a schematic representation of the total Hamiltonian as $H= H_A + H_{A^c} + H_\partial$ may be misleading for non-local theories, since the interactions between $A$ and $B$ need not be confined on the boundary.} Using the Ryu-Takayanagi formula in the context of AdS/CFT correspondence, it can be observed that the regularized entanglement entropy for a $d$-dimensional CFT behaves like thermal entropy: for large enough temperature, the leading order behaviour becomes $S_A \sim V T^{d-1}$, where $V = {\rm vol} \left(\mathbb{R}^{d-1}\right)$. Hence there is no area law at finite temperature at the leading order.

\begin{figure}[!]
\centering
\includegraphics[width=0.7\textwidth]{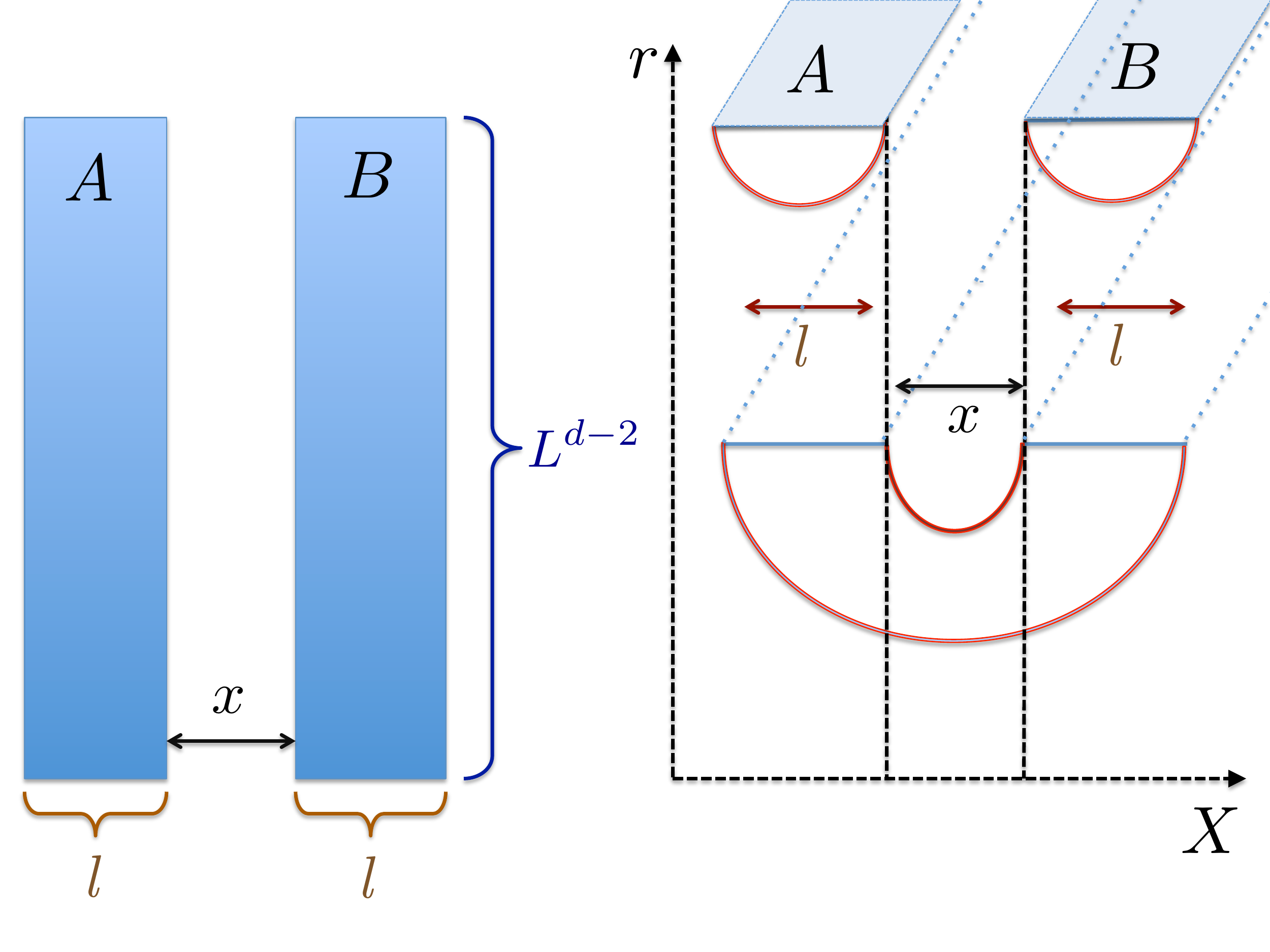}
\caption{The two disjoint sub-systems $A$ and $B$, each of length $l$ along $X$-direction and separated by a distance $x$. The schematic diagram on the right shows the possible candidates for minimal area surfaces which is relevant for computing $S_{A\cup B}$. The choice on top gives $S_{A\cup B} = S_A + S_B = 2 S(l)$; and the choice at the bottom gives $S_{A\cup B} = S(2l+x) + S(x)$. This is also summarized in (\ref{twoch}).}
\label{shape}
\end{figure}

Mutual information is a quantity that is derived from entanglement entropy. The definition of mutual information between two disjoint sub-systems $A$ and $B$ (see fig.~\ref{shape} for example) is given by
\begin{eqnarray} \label{mi}
I(A, B) = S_A + S_B - S_{A \cup B} \ ,
\end{eqnarray}
where $S_A$, $S_B$ and $S_{A\cup B}$ denote entanglement entropy of the region $A$, $B$ and $A\cup B$ respectively with the rest of the system. From the definition, it is clear that mutual information is a finite quantity since the non-universal divergent pieces in the entanglement entropy cancel out. Thus, we do not need to worry about any regularization scheme. Moreover, as showed in \cite{PhysRevLett.100.070502}, given an operator $\O_A$ in the region $A$ and $\O_B$ in the region $B$, mutual information sets an upper bound 
\begin{eqnarray} \label{mi1}
I(A, B) \ge \frac{\left(\langle \O_A \O_B \rangle - \langle \O_A \rangle \langle \O_B \rangle \right)^2}{2 || \O_A ||^2 || \langle \O_B ||^2 }
\end{eqnarray}
and thus measures the total correlation between the two sub-systems: including both classical and quantum correlations. Furthermore, it was shown in \cite{PhysRevLett.100.070502} that mutual information follows an area law even at finite temperature.

In the context of AdS/CFT, or holography, some intriguing features can already be conceived. Let us imagine two disjoint sub-systems $A$ and $B$, each of ``rectangular" shape with one dimension of length $l$ and the other as $L^{d-2}$, are separated by a distance $x$ along one of the spatial directions of a given CFT. This is schematically shown in fig.~\ref{shape}.
One can easily follow the Ryu-Takayanagi formula to compute the entanglement entropy of the individual sub-systems $A$ and $B$. The computation of $S_{A\cup B}$ is more interesting: in this case, there are multiple choices of minimal area surfaces, which are schematically shown in fig.~\ref{shape}.\footnote{Actually, we have not shown yet another possibility where the two minimal area surfaces cross each other. However, this will always have a larger area than the two choices shown in fig.~\ref{shape}.} Depending on the ratio $x/l$,
\begin{eqnarray} \label{twoch}
S_{A \cup B} & = &  S(2l+x) + S(x) \, \, \, \, {\rm for \, \, ``small"} \, \, \, \,  x/l \ , \nonumber\\
                        & = &   2 S(l) \, \, \, \, {\rm for \, \, ``large"} \, \, \, \,  x/l  \ .
\end{eqnarray}
Here $S(y)$ denotes the area of a minimal surface whose boundary has a length of dimension $y$. Thus, in the latter case, we will have $I(A,B) = 0$ identically above a certain value for $x/l$\cite{Headrick:2010zt}.

To summarize, mutual information has an intriguing feature for such systems\cite{Headrick:2010zt}:
\begin{eqnarray} \label{transition}
I(A,B) \not  = && 0 \ , \quad x/l \le a_d \ ,  \nonumber\\
                   =  && 0 \ , \quad x/l > a_d \ .
\end{eqnarray}
Thus, mutual information undergoes a first order phase transition at $x/l = a_d$, where $a_d$ is a number that depends on the dimension of the CFT. By virtue of the relation in (\ref{mi1}), eqn (\ref{transition}) implies that for $x/l > a_d$, the two sub-systems $A$ and $B$ completely disentangle.\footnote{Note that $\rho_{A\cup B} = \rho_A \otimes \rho_B$ implies that $I(A,B) =0$ and {\it vice versa}. We thank Matt Headrick for a correction on this point.} Similar phenomenon persists at finite temperature also. In the limit $l\to\infty$, the disentangling transition takes place as a function of temperature
\begin{eqnarray} \label{transitionT}
I(A,B) & \not = & 0 \ , \quad x T \le b_d \ ,\nonumber\\
& = & 0  \quad x T > b_d \ ,
\end{eqnarray}
where $b_d$ is a constant and $T$ denotes the backgrounds temperature. See \cite{MolinaVilaplana:2011xt} for a related work in AdS$_3$-BTZ black hole background. Our goal here will be to study this disentangling transition for a class of conformal (or scale invariant) large $N$ gauge theories in a general dimension within the context of holography. We will make use of the analytical techniques developed in \cite{Fischler:2012ca} and also use numerical methods to explore the regime of parameters where this disentangling transition takes place in the $(x/l)$ vs $(T x)$ plane.

Before proceeding further, let us offer some more comments. For relativistic CFTs, the area law for mutual information at finite temperature along with dimensional analysis suggests that 
\begin{eqnarray} \label{migen}
I(A,B) = \left(\frac{L}{l} \right)^{d-2} F \left( x/l, x T\right) \ , 
\end{eqnarray}
where $F(x/l, xT)$ is some function that depends on the CFT. At vanishing temperature, we recover the well-known form\cite{Swingle:2010jz}. The two regimes where we are able to obtain analytical results are $lT \ll 1$, $x T \ll 1$ and $lT \gg 1$, $xT \ll 1$ respectively. For small temperature, {\it i.e.}~when both $lT \ll 1$, $x T \ll 1$, we can make a formal expansion of form
\begin{eqnarray} \label{migen1}
F \left(x/l, xT \right) = \sum_{i} (xT)^i g_{i} (x/l) \ . 
\end{eqnarray}
In the limit $lT \gg 1$ but $x T \ll 1$, we can make the following expansion
\begin{eqnarray} \label{migen2}
F\left(x/l, xT \right) = \left(l T\right)^{d-2} \sum_\alpha (xT)^\alpha \tilde{g}_\alpha(lT)  \ .
\end{eqnarray}
where $g_{i}(x/l)$ and $\tilde{g}_\alpha(lT)$ are hitherto undetermined functions that depend on the underlying theory. We will find that generally $i \ge 0$, but $\alpha$ can range over positive and negative numbers. For example, in (\ref{migen2}) as $xT \to 0$, mutual information acquires a divergent piece: $I(A,B) \sim (L/x)^{d-2}$, which is in accord with the results obtained in \cite{Swingle:2010jz}. Finally, in the regime where both $lT \gg1$ and $xT\gg 1$, mutual information vanishes identically. In the next sections, we will discuss some generic examples in the light of equations (\ref{migen1}) and (\ref{migen2}).\footnote{Note that, here we are excluding the possibility of any logarithmic term. In general, such logarithmic contributions can arise; see {\it e.g.}~the example of $(1+1)$-dim CFT and the special case of hyperscaling-violating background in later sections.} Also note that, for non-relativistic scale-invariant theories, equations (\ref{migen1}) and (\ref{migen2}) will have similar forms with $T \to T^{1/z}$, where $z$ denotes the dynamical exponent of the theory.

Before concluding this section, let us also comment on a general result that we will discuss in the subsequent sections. Clearly, both the expansions alluded to in (\ref{migen1}) and (\ref{migen2}) correspond to low temperature with respect to the separation scale, {\it i.e.}~when $x T \ll 1$. However, as we will demonstrate, the two regimes of low and high temperature with respect to the system sizes, {\it i.e.}~for $lT \ll 1$ or $lT \gg 1$ contain distinct physics. It is particularly interesting to consider the case $lT \gg 1$. In this regime, the entanglement entropy of either sub-system $A$ or sub-system $B$ can be schematically given by (see {\it e.g.}~\cite{Fischler:2012ca} or equations (\ref{highee}) and (\ref{higheelif}))
\begin{eqnarray} \label{eeT}
S_{A} = S_B = S_{\rm div} + S_{\rm thermal} + S_{\rm finite} + S_{\rm corr} \ ,
\end{eqnarray}
where $S_{\rm div}$ denotes the divergent piece that typically follows the area law, $S_{\rm thermal}$ denotes the purely thermal entropy that goes as the volume, $S_{\rm finite}$ denotes the next leading order contribution that also follows an area law and finally $S_{\rm corr}$ denotes corrections suppressed by exponentials of $(lT)$. In this limit, mutual information behaves in the following manner:
\begin{eqnarray} \label{miT}
\left. I(A,B) \right |_{x \to 0} = I_{\rm div} + S_{\rm finite} + I_{\rm corr} \ ,
\end{eqnarray}
where $I_{\rm div}$ is the divergent piece that emerges in the limit $x\to 0$ and $I_{\rm div} = S_{\rm div}$ similar to what is observed in \cite{Swingle:2010jz} and $I_{\rm corr}$ are correction terms in powers of $(xT)$ and $e^{- lT}$. From (\ref{eeT}) and (\ref{miT}), we see that apart from the diverging piece as $x\to 0$, mutual information does coincide with the thermal-part-subtracted entanglement entropy at the leading order. Thus, it truly measures quantum entanglement by discarding the volume-worth thermal contribution in the entanglement entropy.

There are perhaps a couple of non-trivialities associated with this observation: First, note that {\it a priori} there is no reason for the sub-leading terms of entanglement entropy to follow an area law in the large temperature regime. This behaviour which was rigorously obtained in \cite{Fischler:2012ca}, however, is very crucial for the above relation to be true. Second, there is a precise match between the numerical factors as well.

\section{Mutual information in relativistic CFTs}

Let us begin by considering a class of large N gauge theories in $d$-dimensions whose dual is given by an asymptotically AdS$_{d+1}$-background. Finite temperature in introduced by having a black hole in the bulk spacetime. The generic bulk spacetime is given by the AdS-Schwarzschild metric of the form
\begin{eqnarray}
ds^2 = - \frac{r^2}{R^2} f(r) dt^2 + \frac{r^2}{R^2} d\vec{x}^2 + \frac{R^2}{r^2} \frac{dr^2 }{f(r)} \ , \quad f(r) = 1 - \frac{r_H^d}{r^d} \ , 
\end{eqnarray}
where $r_H$ is the location of the black hole horizon, $R$ is the AdS radius, $\vec{x}$ is a $(d-1)$-dimensional vector and the boundary of the spacetime is located at $r\to\infty$. The temperature of the background is obtained by Euclideanizing the time direction and periodically compactifying it on a circle. The inverse period of this Euclidean time direction then gives the temperature as:
\begin{eqnarray} \label{temp}
T = \frac{r_H d}{4 \pi R^2} \ .
\end{eqnarray}
In what follows, we will set $R=1$.

To obtain mutual information for an arrangement schematically shown in fig.~\ref{shape}, we specify the strip by
\begin{eqnarray}
X \equiv x^1 \in \left[ - \frac{l}{2}, \frac{l}{2}\right] \ , \quad x^i = \in \left[ - \frac{L}{2}, \frac{L}{2}\right] \ , \quad i = 2, \ldots, d-2 \ .
\end{eqnarray}
with $L \rightarrow \infty$. Extremal surface is translationally invariant along $x^i, i=2,...,d-2$ and the profile of the surface in the bulk is $X(r)$. Area of this surface is given by
\begin{equation} \label{areagen}
A= L^{d-2}\int dr r^{d-2}\sqrt{r^2 X'^2+ \frac{1}{r^2\left(1-\frac{r_H^d}{r^d}\right)}}.
\end{equation}
This action leads to the equation of motion
\begin{align} \label{eomgen}
\frac{dX}{dr}= \pm \frac{r_c^{d-1}}{ r^{d+1} \sqrt{\left(1- \frac{r_c^{2d-2}}{r^{2d-2}}\right)\left(1-\frac{r_H^d}{r^d}\right)}},
\end{align}
where, $r_c$ is an integral of motion and $r=r_c$ represents the point of closest approach of the extremal surface. Such surfaces have two branches, joined smoothly at $(r=r_c, X=0)$ and $r_c$ can be determined using the boundary conditions:
\begin{equation}
X(\infty)=\pm\frac{l}{2} \ ,
\end{equation}
which leads to
\begin{align} \label{lengen}
\frac{l}{2}=&\int_{r_c}^{\infty}\frac{r_c^{d-1} dr}{r^{d+1} \sqrt{\left(1- \frac{r_c^{2d-2}}{r^{2d-2}}\right)}} \left(1-\frac{r_H^d}{r^d}\right)^{-1/2}\nonumber\\
=& \frac{1}{r_c}\int_{0}^{1}\frac{u^{d-1} du}{ \sqrt{1- u^{2d-2}}} \left(1-\frac{r_H^d}{r_c^d}u^d\right)^{-1/2}.
\end{align}
So far, we have kept our discussion for general $d$.

\subsection{Special case: $d=2$}

Let us now focus on $d=2$. In this case, it is possible to evaluate the integrals in (\ref{lengen}) and (\ref{areagen}) in closed forms. This eventually leads to the following expression for entanglement entropy:
\begin{eqnarray} \label{ee2}
S_A = \frac{c}{3} \log\left[ \frac{\beta}{\pi\epsilon} \sinh\left(\frac{\pi l}{\beta}\right)\right] \ , \quad \beta = \frac{1}{T} \ , \quad c = \frac{3}{2 G_N^{(2+1)}} \ . 
\end{eqnarray}
Using the above expressions, the definitions of entanglement entropy and mutual information in (\ref{eedef}) and (\ref{mi}) respectively, we obtain
\begin{eqnarray} \label{mid2}
I(A,B) = \frac{c}{3} \log \left[ \frac{\left(\sinh(\pi l T) \right)^2}{\sinh(\pi xT) \sinh (\pi (2l + x) T)}\right] \ ,
\end{eqnarray}
In the limit $lT \ll 1$ and $xT \ll 1$, we get
\begin{eqnarray}
I (A, B) = \frac{c}{3} \left[\log\left(\frac{l^2}{x (2l+x)}\right) - 	\frac{1}{3} \pi^2 T^2 \left(l + x\right)^2 + \ldots \right] \ ,
\end{eqnarray}
where the first term in the square bracket is just the zero temperature mutual information. In view of (\ref{migen}), we observe that there is no linear term in $T$. We also observe that finite temperature reduces mutual information and therefore promotes disentangling between the two sub-systems.

On the other hand, in the regime $lT \gg 1$ and $xT \ll 1$, we get
\begin{eqnarray} \label{mid2high}
I(A,B) = \frac{c}{3} \left[- \log \left(2\pi x T\right) -  \left(\pi x T\right) + \log\left(\tanh(\pi l T) \right) \ldots \right] \ ,
\end{eqnarray}
where the contributions in $(lT)$ are exponentially suppressed. It is now easy to check that, in the large temperature regime, {\it i.e.}~$l T \gg 1$, the entanglement entropy takes the form
\begin{eqnarray} \label{eed2high}
S_A = S_{\rm div} + \frac{c}{3} \log\left(\sinh (\pi l T)\right) + \ldots \ .
\end{eqnarray}
In the limit $x \to 0$, defining $\epsilon = x/2$ we get that the large temperature expansion of mutual information given in (\ref{mid2high}) coincides exactly with the leading order large temperature expansion of the entanglement entropy given in (\ref{eed2high}). The mismatch is suppressed in exponentials of $(lT)$. This is an example of what we discussed in equations (\ref{eeT}) and (\ref{miT}).

\begin{figure}[!]
\centering
\includegraphics[width=0.8\textwidth]{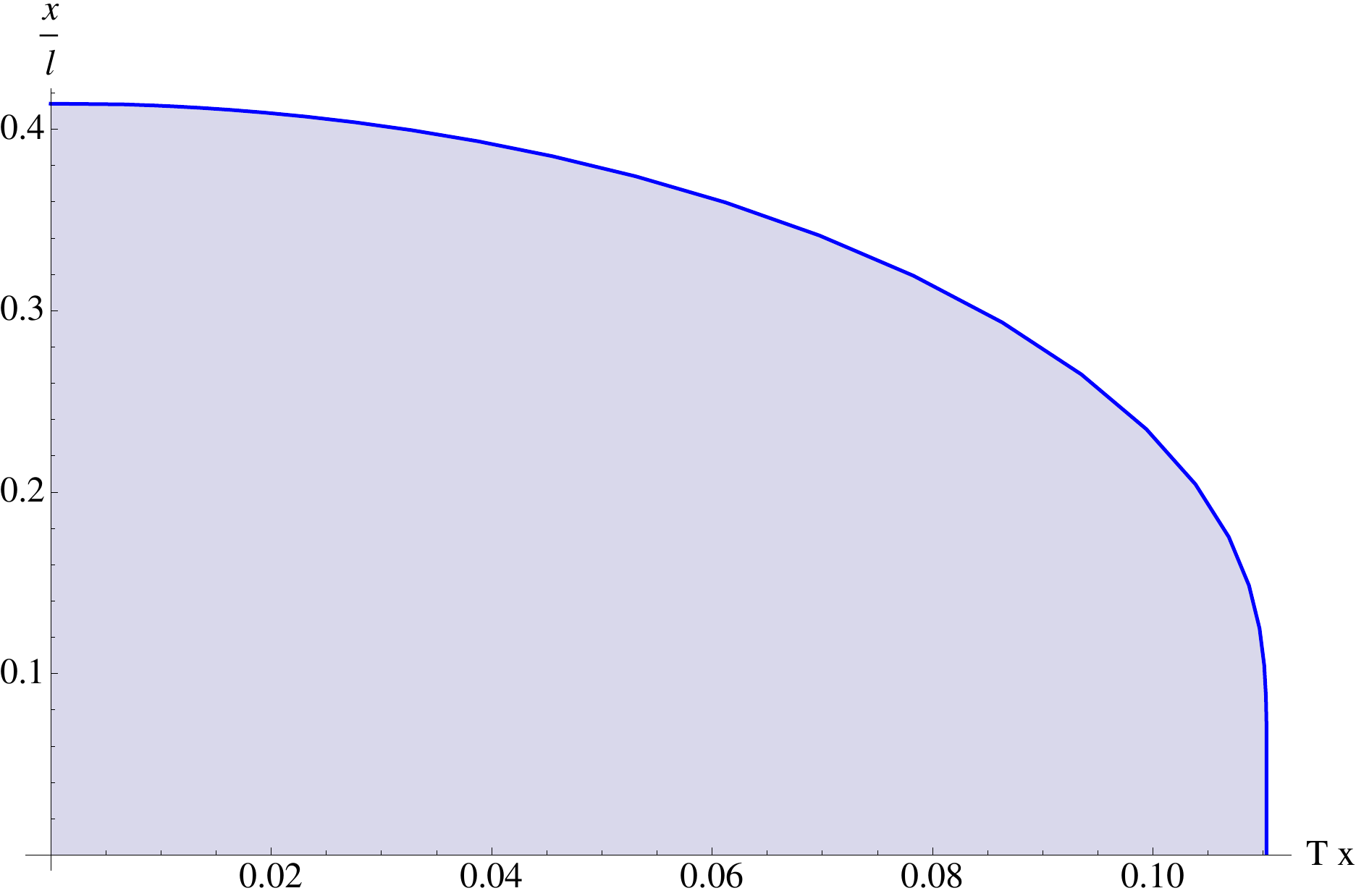}
\caption{2-dimensional parameter space for the (1+1)-dimensional boundary theory. The mutual informational is non-zero only in the blue shaded region.}
\label{2dmi}
\end{figure}

We have pictorially shown a ``phase diagram" in fig.~\ref{2dmi} corresponding to either $I(A,B) \not =0$ or the $I(A,B) = 0$ phase. The blue-shaded region represents the regime of parameters where there is non-vanishing correlation between the two sub-systems. From this phase diagram it is evident that increasing temperature does indeed disentangle the two sub-systems and entanglement reduces monotonically for increasing temperature. In the gravitational dual, increasing temperature implies that the extremal surface probes deeper in the background. Hence, as the two sub-systems keep disentangling, the ``emergent" AdS radial direction becomes more pronounced. Our results are in agreement with earlier work in \cite{MolinaVilaplana:2011xt}.
%

\subsection{General case: $d>2$}

We now move on to discussing the general case of $d>2$. In this case, it is not possible to evaluate the integrals in (\ref{areagen}) and (\ref{eomgen}) in closed forms. We will use the approximation scheme outlined in \cite{Fischler:2012ca}. Much of the relevant details have been relegated to appendix A. Here we will discuss the final results. For the discussions in this section, we will define
\begin{eqnarray} \label{cd}
c = \frac{R^{d-1}}{4 G_N^{(d+1)}} \ , 
\end{eqnarray}
where $G_N^{(d+1)}$ is the Newton's constant in $(d+1)$-dimensional bulk theory. We will set $R=1$.

\subsubsection{Mutual information: $T=0$}

At zero temperature, the mutual information is given by,
\begin{eqnarray} \label{miT0}
I(A,B) & = & c \, \S_0 L^{d-2}\left[\frac{2}{l^{d-2}}-\frac{1}{x^{d-2}}- \frac{1}{(2l+x)^{d-2}}\right] \ , \quad x/l \le a_d \ , \\
           & = & 0 \ , \quad x/l >a_d \ ,
\end{eqnarray}
where $\S_0$ is a constant (defined in (\ref{S0})) with a negative sign, $c$ is defined in (\ref{cd}), $a_d$ is a constant depending on the dimension of the dual CFT. In fig.~\ref{temptran} we have shown how this constant depends on the dimension $d$. It is clear that for a given $x/l$, increasing dimension makes it more difficult to disentangle the two sub-systems. This is intuitively expected since the higher the dimension, the more the ``area" becomes resulting in larger entanglement.

\subsubsection{Finite temperature: $T\ll \frac{1}{l},\frac{1}{x}$}

In this limit, when the mutual information is non-zero, it is given by,
\begin{align}
I(A,B)= & c \, \S_0 L^{d-2}\left[\frac{2}{l^{d-2}}\left(1+ \S_1\left( \frac{4\pi T l}{d}\right)^d\right)-\frac{1}{x^{d-2}}\left(1+ \S_1\left( \frac{4\pi T x}{d}\right)^d\right)\right.\nonumber\\
&\left.- \frac{1}{(2l+x)^{d-2}}\left(1+ \S_1\left( \frac{4\pi T (2l+x)}{d}\right)^d\right)\right]\\
=&I(A,B)|_{T=0}-2 c \,   \S_0\S_1 \left( \frac{4\pi  }{d}\right)^d L^{d-2}~ T^d~ \left(l-x\right)^2 \ .
\end{align}
Here $\S_1$ is a constant (defined in (\ref{S1})) with negative sign. In this case, the finite temperature correction obeys an area law as generally proved in \cite{PhysRevLett.100.070502}. Once again we observe that introducing finite temperature decreases mutual information.

\subsubsection{Finite temperature: $\frac{1}{l}\ll T\ll \frac{1}{x}$}

\begin{figure}[!ht]
\begin{center}
\subfigure[] {\includegraphics[angle=0,
width=0.45\textwidth]{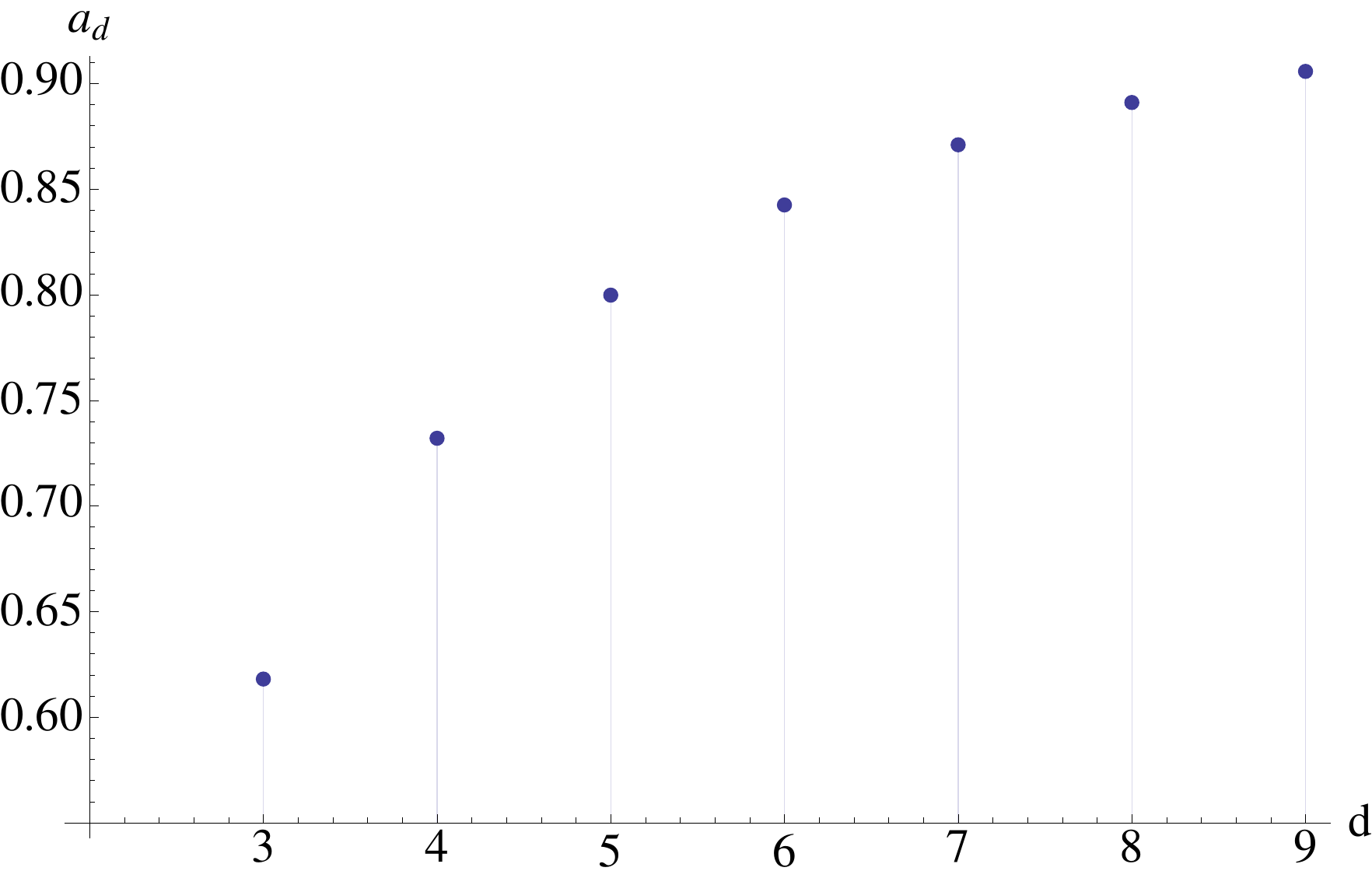} }
 \subfigure[] {\includegraphics[angle=0,
width=0.45\textwidth]{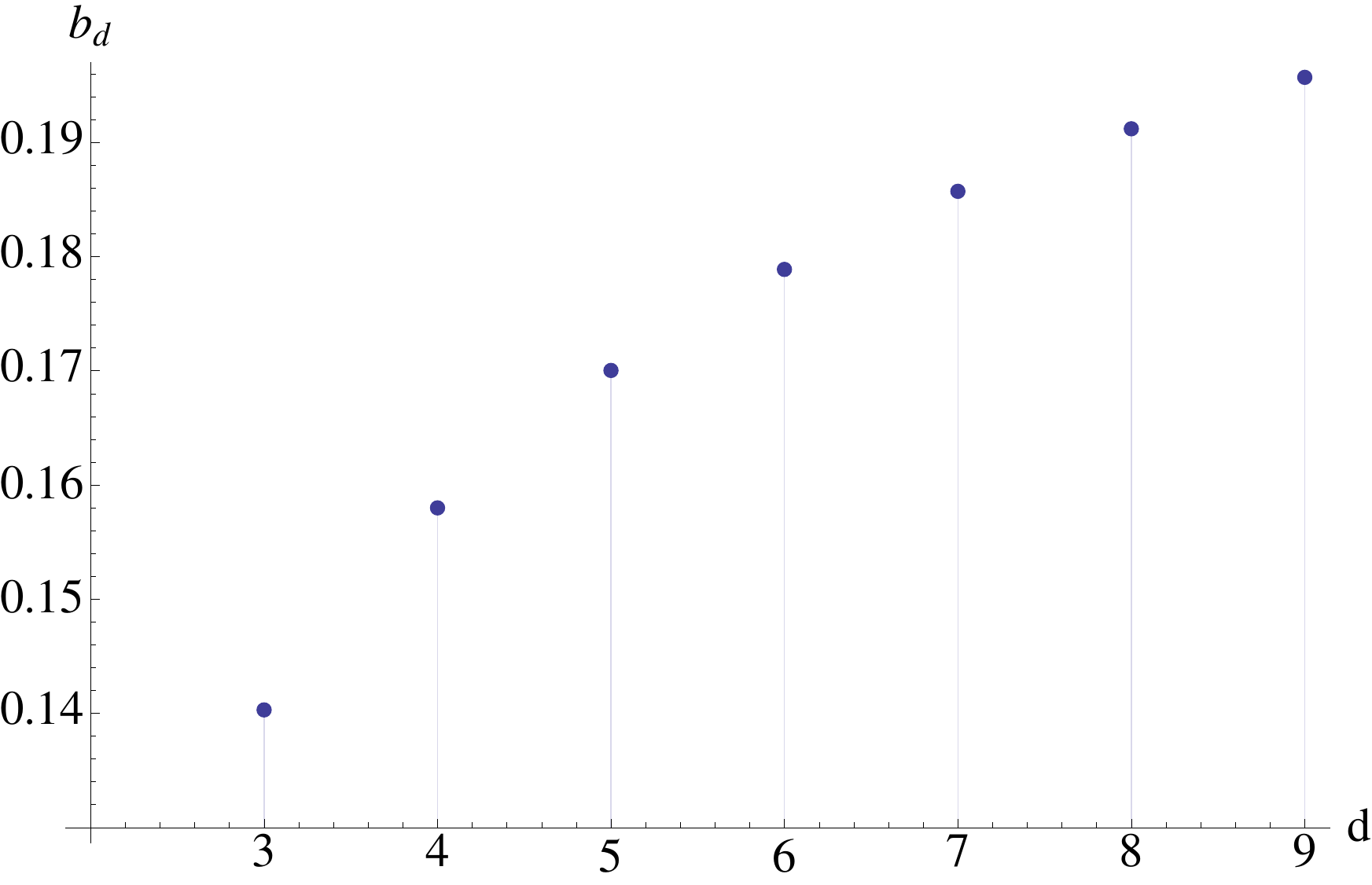} }
\caption{\small The left panel: the dependence of $a_d$, as defined in (\ref{transition}), with respect to $d$. The right panel: the dependence of $b_d$, as defined in (\ref{transitionT}), with respect to $d$. The solid dots represent the corresponding value of $a_d$ or $b_d$ beyond which mutual information vanishes.}
\label{temptran}
\end{center}
\end{figure}

In this limit, when the mutual information is non-zero, it is given by,
\begin{align} \label{midgen}
I(A,B)=  c \,  L^{d-2}T^{d-2}&\left[-\S_0 \frac{1}{(xT)^{d-2}}+\left( \frac{4\pi  }{d}\right)^{d-2}\S_{\rm high}\right.\nonumber\\
&\left.-\left( \frac{4\pi }{d}\right)^{d-1} T x-\S_0\S_1 \left( \frac{4\pi  }{d}\right)^d T^2 x^2\right] \ .
\end{align}
Comparing the expression in (\ref{midgen}) and (\ref{highee}), we again observe that mutual information indeed captures the true entanglement part of the entanglement entropy by getting rid of the thermal contribution and this is a precise result including all numerical factors. It is  an example of the generic observation mention in (\ref{eeT}) and (\ref{miT}). From the above expression, it is possible to find an upper bound on $(xT)\equiv b_d$, above which $I(A,B)$ is always zero. We have shown the dependence of $b_d$ as a function of $d$ in fig.~\ref{temptran}. Once again we observe that increasing dimension increases $b_d$, which is intuitively expected. 
%

\subsubsection{Large temperature: $T\gg \frac{1}{x}$}

\begin{figure}[!]
\centering
\includegraphics[width=0.8\textwidth]{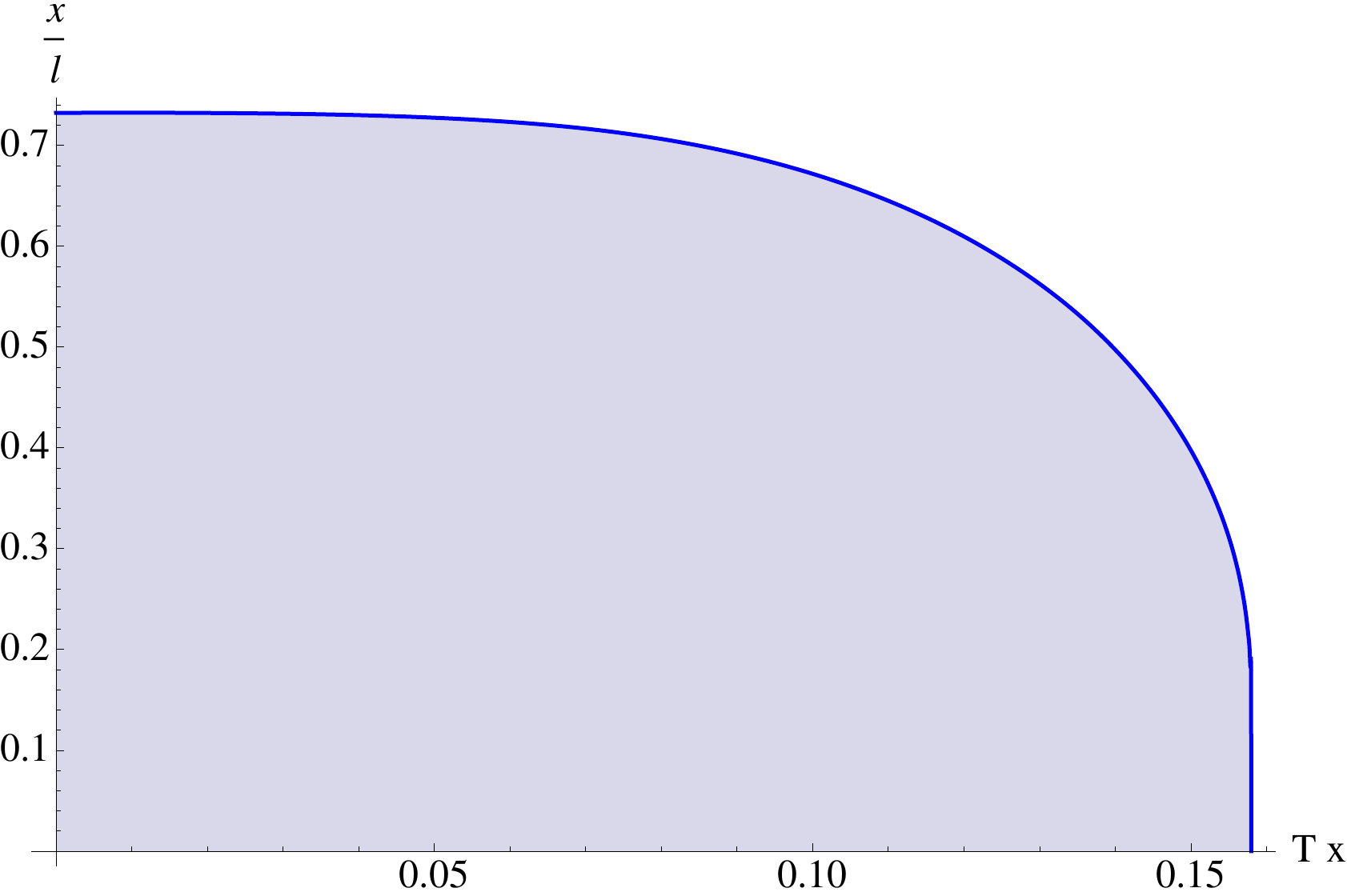}
\caption{2-dimensional parameter space for the (3+1)-dimensional boundary theory. The mutual informational is non-zero only in the blue shaded region. The corresponding parameter space looks qualitatively similar for general $d$, thus we have showed one representative example here.}
\label{4dmi}
\end{figure}

In this limit, the two sub-systems are completely disentangled and mutual information is identically zero. The corresponding ``phase diagram" is shown in fig.~\ref{4dmi}, where the shaded region corresponds to $I(A,B) \not = 0$ and $I(A,B) = 0$ everywhere outside.  
%

\section{Other backgrounds}

We will now consider generic examples of scale-invariant (but not conformal) theories, which are known to have gravity dual descriptions. Such field theories with gravity duals, assuming they exist, are non-relativistic. Examples include the so called Lifshitz geometry introduced in \cite{Kachru:2008yh}; and more recently the background with hyperscaling violation in \cite{Huijse:2011ef}. Note that, in both \cite{Kachru:2008yh} and \cite{Huijse:2011ef} the approach is {\it phenomenological} or the so called {\it bottom-up}, {\it i.e.}~the existence of a dual field theory with the right symmetry properties is postulated {\it ab initio} without directly making connection to a more rigorous string or brane-construction. A lot of progress have been made to embed such effective gravity descriptions in ten or eleven dimensional supergravity, see {\it e.g.}~\cite{Balasubramanian:2010uk, Donos:2010tu} and \cite{Dong:2012se, Narayan:2012hk} respectively. By now there is a vast literature on such embeddings, and emboldened by these results we will work with an effective description without explicitly referring to the precise details of the dual field theory and also assume that the Ryu-Takayanagi proposal holds.

\subsection{Lifshitz background}

Let us discuss the Lifshitz background first. In this case, the background metric is invariant under the following scale transformation:
\begin{eqnarray}
t \to \lambda^z t \ , \quad x \to \lambda x \ , \quad r \to \lambda r \ ,
\end{eqnarray}
where $\lambda$ is a real number and $r$ is the radial coordinate, in which the boundary is located at $r\to 0$. Such backgrounds are typically obtained from Einstein gravity with a negative cosmological constant and some matter field, such as a massive vector field or a scalar field. An analytic finite temperature Lifshitz background in $(3+1)$-dimensions is obtained in \cite{Balasubramanian:2009rx}, and is given by
\begin{eqnarray} \label{Lifshitz}
&& ds^2 = R^2\left(- f \frac{dt^2}{r^{2z}} + \frac{d\vec{x}^2}{r^2} + \frac{dr^2}{f r^2} \right)\ , \quad f = 1- \frac{r^2}{r_H^2} \ , \\
&& \phi = - \frac{1}{2} \log\left( 1 + \frac{r^2}{r_H^2} \right) \ , \quad A = \frac{f}{r^2} dt \ ,
\end{eqnarray}
where $\phi$ is the dilaton field and $A$ is a massive vector field and the dynamical exponent $z=2$. For our purposes, it is only the background metric that will be relevant. The temperature in the dual field theory is given by the Hawking temperature of the black hole
\begin{equation}
T=\frac{R}{2\pi r_H^2} \ .
\end{equation}

Now the details of the calculations for entanglement entropy and subsequently mutual information will proceed as before. The relevant details are provided in appendix B. Here we will discuss the final results. At zero temperature, the mutual information is given by
\begin{equation} \label{lifmi}
I(A,B)=-c \, \L_0 L\left[\frac{2}{l}-\frac{1}{x}- \frac{1}{(2l+x)}\right] \ , \quad c = \frac{R^2}{4G_N^{(4)}} \ ,
\end{equation}
for $x/l\le 0.618$. Here $\L_0$ is a numerical constant given in (\ref{l0l1}). Note that the above formula matches exactly with the zero temperature mutual information for $d=3$ obtained in (\ref{miT0}). This is expected since at zero temperature, the dynamical exponent of the background does not enter in the computations. Thus it is not possible to distinguish between a relativistic CFT and a scale-invariant on-relativistic field theory by looking at the behaviour of the mutual information. 

In the intermediate temperature range: $\sqrt{T/R}\ll \frac{1}{l},\frac{1}{x}$, we get
\begin{equation}
I(A,B) =I(A,B)|_{T=0} - 2 c \,   \L_0\L_1 \frac{ L~ T}{R} x + \ldots \ ,
\end{equation}
where $\L_1$ is a numerical constant given in (\ref{l0l1}). In this case, in addition to the familiar area law, we do observe a linear correction in temperature as a small temperature is introduced. As before, introducing temperature decreases mutual information.

\begin{figure}[!]
\centering
\includegraphics[width=0.7\textwidth]{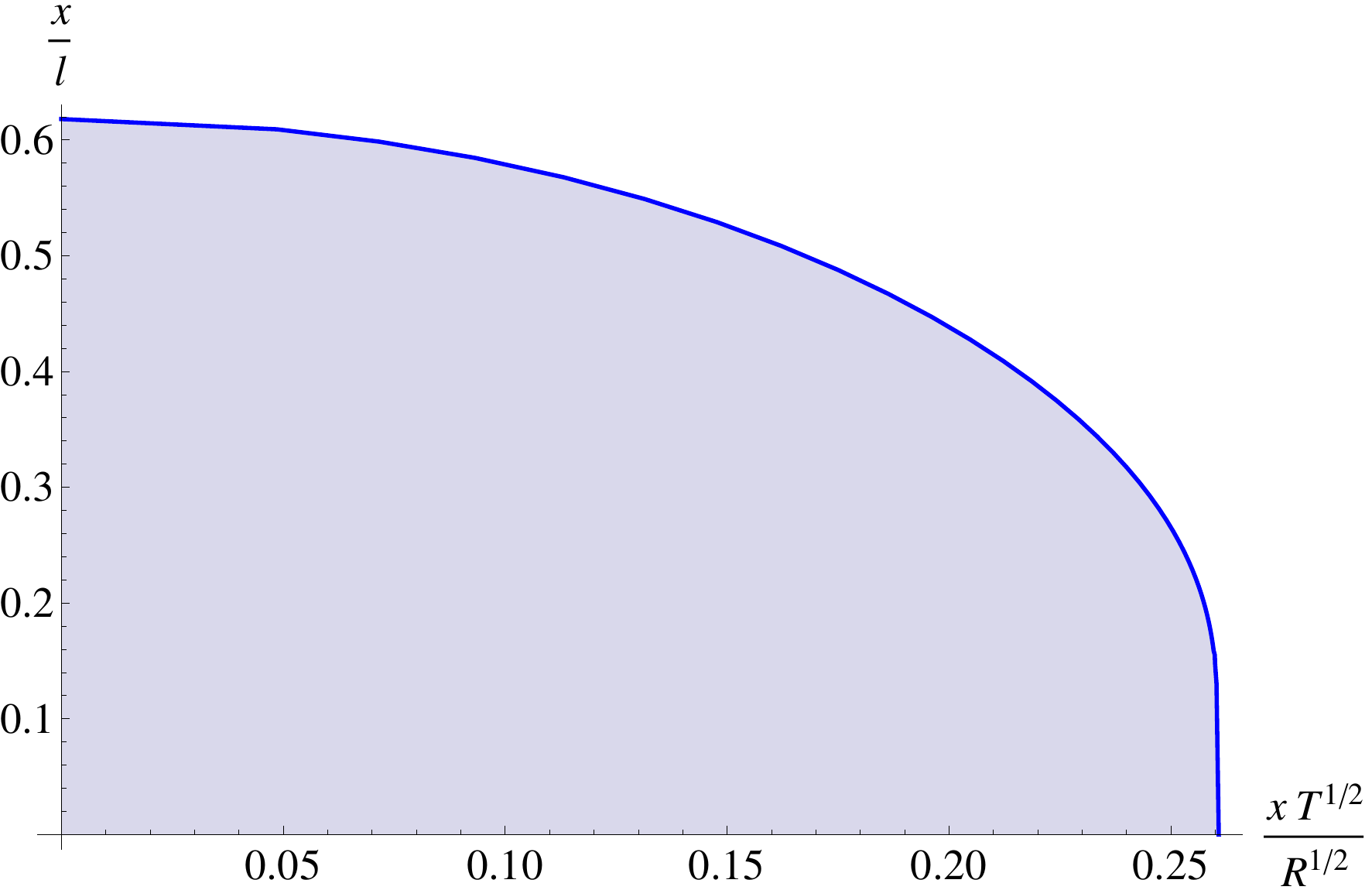}
\caption{\small $2$-dimensional parameter space for a scale-invariant $(2+1)$-dimensional field theory with Lifshitz scaling. The dynamical exponent is $z=2$. Mutual information is non-zero in the shaded region.}
\label{lifpd}
\end{figure}

In the limit $\frac{1}{l}\ll \sqrt{T/R}\ll \frac{1}{x}$, mutual information is obtained to be:
\begin{eqnarray} \label{mihighlif}
I(A,B) & = & c \, L \sqrt{\frac{T}{R}}\left[\L_0 \sqrt{\frac{R}{T}}\frac{1}{x}-\L_{\rm high} -\sqrt{\frac{T}{R}}x(2\pi+\L_0\L_1)\right] \ , \quad  \sqrt{\frac{T}{R}}x \le 0.261 \ , \\
           & = & 0  \ , \quad \sqrt{\frac{T}{R}}x > 0.261 \ .
\end{eqnarray}
Here $\L_{\rm high} = 2.671$ is just a numerical constant. Comparing (\ref{mihighlif}) with (\ref{higheelif}), we find that in this regime mutual information indeed coincides with the thermal-part-subtracted entanglement entropy. Finally for $\sqrt{T/R}\gg \frac{1}{x}$, $I(A,B) = 0$ identically. The corresponding $2$-dimensional ``phase diagram" is shown in fig.~\ref{lifpd}, where the shaded region corresponds to $I(A,B) \not = 0$ and it vanishes everywhere else.
%

\subsection{Hyperscaling-violating background}

A more general background with hyperscaling violation was proposed in \cite{Huijse:2011ef}. In this case, the metric is covariant under the scale transformation and has the following property:
\begin{eqnarray}
t \to \lambda^z t \ , \quad r \to \lambda r \ , \quad x \to \lambda x \ , \quad ds^2 \to \lambda^{2\theta/(d-1)} ds^2 \ ,
\end{eqnarray}
where $\theta$ is known as the hyperscaling violation exponent. In the presence of a black hole, the metric takes the following form\cite{Alishahiha:2012qu}\footnote{Note that in \cite{Alishahiha:2012qu} $d$ denotes the spatial dimensions only. Thus $d_{\rm here} = d_{\rm there} + 1$.}
\begin{eqnarray}
&& ds^2 = r^{2\theta/(d-1)} \left(- f(r) \frac{dt^2}{r^{2z}} + \frac{dr^2}{r^2 f(r)} + \frac{d\vec{x}^2}{r^2} \right) \ , \nonumber\\
&& f(r) = 1 - \left(\frac{r}{r_H}\right)^{\gamma} \ ,
\end{eqnarray}
where $\gamma$ is a real-valued constant which we will keep unspecified for now, $r_H$ is the location of the horizon, $z$ is the dynamical exponent, $\theta$ is the hyperscaling violation exponent and $d$ is the spacetime dimension of the boundary dual theory. We have also set the curvature of the space $R=1$. The advantage of writing the metric in the above fashion is the fact that in the zero temperature limit it becomes conformal to the Lifshitz metric written in (\ref{Lifshitz}). The boundary is located at $r \to 0$. The backgrounds temperature is given by
\begin{equation}
T= \frac{\gamma}{4 \pi r_H^z} \ .  
\end{equation} 
We will consider the case when $d-\theta-2\ge 0$, which typically exhibits an area law for entanglement entropy with the exception of logarithmic violation for $\theta = d - 2$.

\subsubsection{General case: $\theta \not = d-2$}

In the same spirit as before, let us investigate the general case of $\theta \not = d-2$ in various temperature regimes. Some relevant details containing the high temperature and low temperature expansions for entanglement entropy have been relegated in appendix C. \\

\noindent{\bf The case of $T=0$:} At zero temperature, the mutual information for the $d$-dimensional boundary theory (for $d-\theta-2> 0$), is given by
\begin{equation} \label{hsmi}
I(A,B)= c \, \C(\theta,d) L^{d-2}\left[\frac{2}{l^{d-\theta-2}}-\frac{1}{x^{d-\theta-2}}- \frac{1}{(2l+x)^{d-\theta-2}}\right] \ , \quad c = \frac{1}{4G_N^{(d+1)}} \ ,
\end{equation}
where $\C(\theta, d)$ is a numerical constant which is given in appendix C. $I(A,B)$ is zero when $x/l\ge a_{hs}$, where $a_{hs}$ is the solution of the algebraic equation
\begin{equation}
2 a_{hs}^{d-\theta-2}-1- \frac{a_{hs}^{d-\theta-2}}{(2+a_{hs})^{d-\theta-2}}=0 \ .
\end{equation}
\\

\noindent{\bf The case of $l T^{1/z}, x T^{1/z}\ll 1$:} In this limit, when the mutual information is non-zero, it is given by,
\begin{equation}
I(A,B) =I(A,B)|_{T=0}+ c \, h_1 \C(\theta,d) L^{d-2}T^{\frac{\gamma}{z}}  \left[2 l^{\gamma-d+\theta + 2}-x^{\gamma-d+\theta + 2}-(2l+x)^{\gamma-d+\theta + 2}\right] \ .
\end{equation}
Here $h_1$ is a numerical constant, which does not contain any physical information. We do observe the familiar correction term at low temperature. \\

\noindent{\bf The case of $ x T^{1/z}\ll 1, l T^{1/z}\gg1$:} In this limit, when the mutual information is non-zero, it is given by
\begin{equation} \label{mihshigh}
I(A,B) =-c \, L^{d-2} \left[\frac{\C(\theta,d)}{x^{d-\theta-2}} - h_3 T^{\frac{d - \theta - 2}{z} } + \C(\theta,d)h_1 T^{\frac{\gamma}{z}}x^{\gamma-d+\theta+2}+h_2 T^{\frac{d-\theta-1}{z}}x\right] \ ,
\end{equation}
where $h_2$ is a numerical constant. Finally, as before we have $I(A,B) = 0$ identically in the limit $ x T^{1/z}\gg 1$. Comparing equations (\ref{mihshigh}) with (\ref{eehighhs}), we note that mutual information coincides with the thermal-part-subtracted entanglement entropy at large temperature. It can be checked that the corresponding ``phase diagram" looks very similar to the ones analyzed before; hence we do not explicitly provide one here.

\subsubsection{Special case: $\theta = d-2$}

Let us now consider the special case of $\theta = d-2$, where logarithmic violation of the area law shows up. For a similar earlier study, see {\it e.g.}~\cite{Huijse:2011ef}.\\

\noindent{\bf The case of $T = 0$:} At zero temperature, the mutual information for the $d-$dimensional boundary theory (for $d-\theta-2= 0$), is given by
\begin{equation}
I(A,B)=c \, L^{d-2}\ln\left[\frac{l^2}{x(2l+x)}\right] \ ,
\end{equation}
$I(A,B)$ is zero when $x/l\ge 0.414$. This result is identical to the result obtained in (\ref{mid2}) when $T=0$. \\

\noindent{\bf The case of $l T^{1/z}, x T^{1/z}\ll 1$:}  In this limit, when the mutual information is non-zero, it is given by
\begin{equation}
I(A,B) =I(A,B)|_{T=0}+ 2L^{d-2}c \, k_1 T^{\gamma/z}\left[2 l^{\gamma}-x^{\gamma}-(2l+x)^{\gamma}\right] \ ,
\end{equation}
where $k_1$ is a numerical constant. \\

\noindent{\bf The case of $ x T^{1/z}\ll 1, l T^{1/z}\gg1$:} In this limit, when the mutual information is non-zero, it is given by
\begin{equation}
I(A,B) = c L^{d-2} \left[2 \ln \left(\frac{1}{x T^{1/z}}\right)+k_2 - k_3 x T^{1/z}- 2k_1 x^{\gamma} T^{\gamma/z}\right] \ ,
\end{equation}
where $k_2$ and $k_3$ are numerical constants. In this case, irrespective of the value of $\gamma$, mutual information does indeed capture the thermal-part-subtracted entanglement entropy. Finally, in this limit of large temperature, $I(A,B) = 0$ identically. This also results in a similar ``phase diagram".

\section{Conclusions and Outlook}

In this article, we have explored the disentangling transition between two sub-systems by studying mutual information in the context of holography. We have considered a class of large $N$ relativistic gauge theories as well as generic examples of non-relativistic scale-invariant theories. We have found an universal qualitative behaviour in the corresponding ``phase diagram" in $(x/l)$ vs $(xT)$-plane.

There are numerous possibilities that we can consider in future. It will be interesting to explore how the disentangling transition depends on the shape of the sub-systems $A$ and $B$. Intuitively, we expect this transition to be present irrespective of the geometry of the sub-systems, but the precise nature of the transition may depend crucially on it.

In recent years, there have been a lot of developments in understanding and holographically computing a more general notion of entanglement entropy, the so called R\'{e}nyi entropy. See for example \cite{Casini:2011kv, Hung:2011nu}. Subsequently we can define a mutual information that is derived from the R\'{e}nyi entropy. It is an interesting question to explore what physics may be contained in this R\'{e}nyi mutual information as far as the disentangling transition is concerned.

In gravity duals of confining large $N$ gauge theories, it is known that the entanglement entropy itself undergoes a transition\cite{Nishioka:2006gr, Klebanov:2007ws}: this corresponds to having two candidate minimal area surfaces for a given length. Thus it will be extremely interesting to explore the physics of mutual information in such backgrounds, since in addition to the disentangling transition that we have explored here, the transition in the entanglement entropy itself is likely to produce a richer analogue of the ``phase diagram" that we have analyzed here.

In quantum many-body systems, mutual information is emerging as an useful order parameter for certain phase transitions, such as the ones described in \cite{Singh:2011, 2012JSMTE..01..023W}. Within the context of AdS/CFT correspondence or the gauge-gravity duality, examples of various phase transitions are plentiful. Typically such phase transitions are engineered to understand aspects of strongly coupled Quantum Chromodynamics (QCD) or more recently in strongly coupled condensed matter-type systems, see {\it e.g.}~\cite{Hartnoll:2009sz} for a review on some of these. It will be interesting to consider what role mutual information might play in phase transitions that are described within the context of holography.

Finally, let us note that the sharp transition of mutual information is a consequence of large N limit. In this limit, the inequality in (\ref{mi1}) is trivially satisfied since the right hand side is always $1/N$-suppressed\cite{Headrick:2010zt}. At finite $N$, however, mutual information should not vanish identically. Hence, the $1/N$-corrections to the RT formula perhaps do not contain a simple geometric interpretation as an area functional in the bulk geometry.

\section{Acknowledgements}

We are grateful to Matthew Headrick for his insightful comments and discussions that led to this work and also for pointing out a couple of errors in a previous version. This material is based upon work supported by the National Science Foundation under Grant Number PHY-0969020 and by Texas Cosmology Center, which is supported by the College of Natural Sciences and the Department of Astronomy at the University of Texas at Austin and the McDonald Observatory. AK is also supported by a Simons postdoctoral fellowship awarded by the Simons Foundation.

\renewcommand{\theequation}{A.\arabic{equation}}
\setcounter{equation}{0}  
\section*{Appendix A. Low and high temperature expansions}
\addcontentsline{toc}{section}{Appendix A. Low and high temperature expansions}

Here we will recall some of the relevant results that have been discussed in details in \cite{Fischler:2012ca}. Let us make the following expansion of (\ref{lengen})
\begin{align}
l=\frac{2}{ r_c}\sum_{n=0}^\infty\left(\frac{1}{1+nd}\right)\frac{\Gamma\left[\frac{1}{2}+n\right]\Gamma \left[\frac{d (n+1)}{2 (d-1)}\right]}{ \Gamma[1+n]\Gamma \left[\frac{d n+1}{2 (d-1)}\right]}\left(\frac{r_H}{r_c}\right)^{nd} \ , \label{eerc}
\end{align}
which converges for $r_c> r_H$. The area of the extremal surface is given by
\begin{align}
A=2 L^{d-2}\int_{r_c}^{\infty}\frac{r^{d-3} dr}{ \sqrt{\left(1- \frac{r_c^{2d-2}}{r^{2d-2}}\right)}} \left(1-\frac{r_H^d}{r^d}\right)^{-1/2} \ ,
\end{align}
which is a divergent quantity with the divergent piece:
\begin{equation}
A_{\rm div}=\frac{2}{d-2}L^{d-2}{r_b}^{d-2}= \frac{2}{d-2}\left(\frac{L}{a}\right)^{d-2}\qquad d\neq 2 \ . 
\end{equation}
Here $r_b$ corresponds to the ultraviolet cut off $a=1/r_b$ (or a lattice spacing) of the boundary theory.\footnote{We are working with AdS radius $R=1$. Restoring $R$, the lattice spacing is given by $a=\frac{R^2}{r_b}$.}
This is the familiar area law divergence. This area law behavior of the divergent piece is well understood from field theory computations \cite{Bombelli:1986rw, Srednicki:1993im}.

Now, we can do an expansion $(d\neq 2)$ for the finite part of the area
\begin{align}
A_{\rm finite}=&2 L^{d-2}r_c^{d-2}\int_{r_c/r_b}^{1}\frac{ du}{u^{d-1} \sqrt{1- u^{2d-2}}} \left(1-\frac{r_H^d}{r_c^d}u^d\right)^{-1/2}-\frac{2}{d-2}L^{d-2}{r_b}^{d-2}\nonumber\\
=&2 L^{d-2}r_c^{d-2} \left[\frac{\sqrt{\pi } \Gamma \left(-\frac{d-2}{2 (d-1)}\right)}{2 (d-1) \Gamma \left(\frac{1}{2 (d-1)}\right)}+ \sum_{n=1}^\infty\left(\frac{1}{2(d-1)}\right)\frac{\Gamma\left[\frac{1}{2}+n\right]\Gamma \left[\frac{d (n-1)+2}{2 d-2}\right]}{ \Gamma[1+n]\Gamma \left[\frac{d n+1}{2 (d-1)}\right]}\left(\frac{r_H}{r_c}\right)^{nd}\right] \ , \label{aren}
\end{align}
which again converges for $r_c>r_H$. Now we can solve equation (\ref{eerc}) for $r_c$ and then we can calculate area by using equation (\ref{aren}); entanglement entropy of the rectangular strip can subsequently be computed using (\ref{eedef}). We can then extract low and high temperature behavior of the entanglement entropy from equations (\ref{eerc}, \ref{aren}).\\

\noindent {\bf Low temperature regime:}\\

The low temperature regime is characterized by having $Tl\ll 1$, or equivalently $r_H l \ll 1$. In this case, $r_c\gg r_H$ and the leading contributions to the area come from the near-boundary AdS region. The deviations can be computed to give 
\begin{align}
r_c=\frac{2 \sqrt{\pi}\Gamma\left[\frac{d}{2(d-1)}\right]}{l~\Gamma\left[\frac{1}{2(d-1)}\right]} \left[1+ \frac{1}{2(d+1)}\frac{2^{\frac{1}{d-1}-d} \Gamma \left(1+\frac{1}{2 (d-1)}\right) \Gamma \left(\frac{1}{2 (d-1)}\right)^{d+1}}{\pi^{\frac{d+1}{2}}  \Gamma \left(\frac{1}{2}+\frac{1}{d-1}\right)\Gamma\left(\frac{d}{2(d-1)}\right)^d}\left(r_H l\right)^d+\O\left(r_H l\right)^{2d}\right]
\end{align}
Now using equation (\ref{aren}), at first order in $(r_H l)^d$, we get 
\begin{align}
A_{\rm finite}= \S_0 \left(\frac{L}{l}\right)^{d-2}\left[1+ \S_1 (r_H l)^d+ \O(r_H l)^{2d}\right] \ , 
\end{align}
where, 
\begin{align} 
\S_0=& \frac{2^{d-2} \pi ^{\frac{d-1}{2}} \Gamma \left(-\frac{d-2}{2 (d-1)}\right) }{(d-1) \Gamma \left(\frac{1}{2 (d-1)}\right)} \left(\frac{\Gamma \left(\frac{d}{2 (d-1)}\right)}{\Gamma \left(\frac{1}{2 (d-1)}\right)}\right)^{d-2} \ , \label{S0} \\
\S_1=& \frac{\Gamma \left(\frac{1}{2 (d-1)}\right)^{d+1}}{\Gamma \left(\frac{d}{2(d-1)}\right)^d\Gamma \left(\frac{1}{2}+\frac{1}{d-1}\right)}2^{-d-1} \pi ^{-\frac{d}{2}} \left(\frac{\Gamma \left(\frac{1}{d-1}\right) }{\Gamma \left(-\frac{d-2}{2 (d-1)}\right)}+\frac{2^{\frac{1}{d-1}} (d-2) \Gamma \left(1+\frac{1}{2 (d-1)}\right) }{\sqrt{\pi } (d+1)}\right) \ . \label{S1}
\end{align}
Therefore, following equation (\ref{eedef}), the entanglement entropy of the rectangular strip for the $d$-dimensional boundary theory at low temperature ($Tl\ll1$) is given by,
\begin{equation}
S_A=c \left[\frac{2}{d-2}\left(\frac{L}{a}\right)^{d-2}+ \S_0 \left(\frac{L}{l}\right)^{d-2}\left\{1+ \S_1 \left(\frac{4 \pi T l}{d}\right)^d+ \O\left(\frac{4 \pi T l}{d}\right)^{2d}\right\}\right] \ ,
\end{equation} 
where $c$ is defined in (\ref{cd}). \\

\noindent{\bf High temperature regime}\\

At high temperature (i.e. $T l \gg 1$, or equivalently $r_H l \gg 1$), using the methods outlined in \cite{Fischler:2012ca} we can evaluate a perturbative expansion. In this case ($r_H l\gg 1$), $r_c$ approaches $r_H$. We will rewrite equation (\ref{aren}) in a way that allows us to take the limit  $r_c\rightarrow r_H$ without encountering any divergence.
\begin{align}
A_{\rm finite}=&2 L^{d-2}r_c^{d-2} \left[\frac{\sqrt{\pi } \Gamma \left(-\frac{d-2}{2 (d-1)}\right)}{2 (d-1) \Gamma \left(\frac{1}{2(d-1)}\right)}\right.\nonumber\\&\left.+ \sum_{n=1}^\infty\frac{1}{1+nd}\left(1+\frac{d-1}{d(n-1)+2}\right)\frac{\Gamma\left[\frac{1}{2}+n\right]\Gamma \left[\frac{d (n+1)}{2 d-2}\right]}{ \Gamma[1+n]\Gamma \left[\frac{d n+1}{2 (d-1)}\right]}\left(\frac{r_H}{r_c}\right)^{nd}\right] \nonumber\\
=&2 L^{d-2}r_c^{d-2} \left[\frac{l r_c}{2}-\frac{\sqrt{\pi }(d-1) \Gamma \left(\frac{d}{2 (d-1)}\right)}{(d-2)\Gamma \left(\frac{1}{2(d-1)}\right)}\right.\nonumber\\
&+\left.\sum_{n=1}^\infty\left(\frac{1}{1+nd}\right)\left(\frac{d-1}{d(n-1)+2}\right)\frac{\Gamma\left[\frac{1}{2}+n\right]\Gamma \left[\frac{d (n+1)}{2 d-2}\right]}{ \Gamma[1+n]\Gamma \left[\frac{d n+1}{2 (d-1)}\right]}\left(\frac{r_H}{r_c}\right)^{nd}\right].\label{highA}
\end{align}
The infinite series in the last equation for large $n$ admits the limit $r_c\rightarrow r_H$. The leading behavior is obtained to be
\begin{align}\label{Shigh}
A_{\rm finite}\approx l L^{d-2}r_H^{d-1}\left[1+ \left(\frac{1}{l r_H}\right)\S_{\rm high}\right] \ , 
\end{align}
where, 
\begin{align}
\S_{\rm high}=&2\left[-\frac{\sqrt{\pi }(d-1) \Gamma \left(\frac{d}{2 (d-1)}\right)}{(d-2)\Gamma \left(\frac{1}{2(d-1)}\right)}+\sum_{n=1}^\infty\left(\frac{1}{1+nd}\right)\left(\frac{d-1}{d(n-1)+2}\right)\frac{\Gamma\left[\frac{1}{2}+n\right]\Gamma \left[\frac{d (n+1)}{2 d-2}\right]}{ \Gamma[1+n]\Gamma \left[\frac{d n+1}{2 (d-1)}\right]}\right] \ .\label{constantshigh}
\end{align}
Hence, the entanglement entropy of the rectangular strip for the $d$-dimensional boundary thoery at high temperature is given by,
\begin{equation}\label{highee}
S_A =  c \left[\frac{2}{d-2}\left(\frac{L}{a}\right)^{d-2}+V \left(\frac{4 \pi T}{d}\right)^{d-1}\left\{1+ \left(\frac{d}{4 \pi T l }\right)\S_{\rm high}\right\} + \ldots \right] \ , 
\end{equation}
where $V=l L^{d-2}$ is the volume of the rectangular strip with AdS radius $R=1$.

\renewcommand{\theequation}{B.\arabic{equation}}
\setcounter{equation}{0}  
\section*{Appendix B. Computations in the Lifshitz background}
\addcontentsline{toc}{section}{Appendix B. Computations in the Lifshitz background}

In this case, the area functional of the surface is given by
\begin{equation}
A= L\int dr \sqrt{X'^2+ \frac{1}{\left(1-\frac{r^2}{r_H^2}\right)}} \ .
\end{equation}
This action leads to the equation of motion
\begin{align}
\frac{dX}{dr}= \pm \frac{r^{2}}{ r_c^{2} \sqrt{\left(1- \frac{r^{4}}{r_c^{4}}\right)\left(1-\frac{r^2}{r_H^2}\right)}} \ ,
\end{align}
where, $r_c$ can be determined from
\begin{equation}
x(\infty)=\pm\frac{l}{2} \ .
\end{equation}
Thus we get
\begin{align}
l=2\int^{r_c}_{0} \frac{r^{2}dr}{ r_c^{2} \sqrt{\left(1- \frac{r^{4}}{r_c^{4}}\right)\left(1-\frac{r^2}{r_H^2}\right)}} = & 2r_c \int_{0}^{1}\frac{u^2 du}{ \sqrt{1- u^{4}}} \left(1-\frac{r_c^2}{r_H^2}u^2\right)^{-1/2}\nonumber\\
=&\frac{r_c}{ 2}\sum_{n=0}^\infty\frac{\Gamma\left[\frac{1}{2}+n\right]\Gamma \left[\frac{n}{2}+\frac{3}{4}\right]}{ \Gamma[1+n]\Gamma \left[\frac{n}{2}+\frac{5}{4}\right]}\left(\frac{r_c}{r_H}\right)^{2n} \ .\label{lifrc}
\end{align}
For any finite temperature  $r_c< r_H$ and the infinite series converges. The area of the extremal surface is given by
\begin{align}
A=2L  \int^{r_c}_{0} \frac{dr}{ r^{2} \sqrt{\left(1- \frac{r^{4}}{r_c^{4}}\right)\left(1-\frac{r^2}{r_H^2}\right)}} \ .
\end{align}
This area is infinite indicating that the entanglement entropy has a divergence. We can do a similar expansion for the area
\begin{align}
A=\frac{2L}{a}+ \frac{L}{r_c}\left[-\frac{(2\pi)^{3/2}}{\Gamma(1/4)^2}+\frac{1}{2} \sum_{n=1}^\infty\frac{\Gamma\left[\frac{1}{2}+n\right]\Gamma \left[\frac{n}{2}-\frac{1}{4}\right]}{ \Gamma[1+n]\Gamma \left[\frac{n}{2}+\frac{1}{4}\right]}\left(\frac{r_c}{r_H}\right)^{2n}\right] \ , \label{lifee}
\end{align} 
where, again $a$ is the ultraviolet cut (or a lattice spacing) of the boundary theory.

At low temperature, $r_c\ll r_H$ and the leading contributions to the area come from the boundary. In this limit,  equation (\ref{lifrc}) can be solved for $r_c$ and at first order, we obtain
\begin{equation}
r_c=\frac{\Gamma(1/4)^2 l}{(2\pi)^{3/2}}\left[1- \frac{\Gamma(1/4)^8}{12 (2\pi)^{5}}\frac{l^2}{r_H^2}+\O\left(\frac{l^4}{r_H^4}\right)\right] \ .
\end{equation}
Now using equation (\ref{lifee}), the entanglement entropy of the rectangular strip for the boundary theory at low temperature ($l\sqrt{T/R}\ll1$) is given by
\begin{equation}
S_A=c \, \left[\frac{2L}{a} -\L_0\left(\frac{L}{l}\right)\left\{1-\L_1\frac{Tl^2}{R} + \ldots \right\}\right] \ ,
\end{equation}
where, $\L_0, \L_1$ are numerical constants given by 
\begin{align} \label{l0l1}
\L_0= \frac{(2\pi)^{3}}{\Gamma(1/4)^4} \ , \qquad \L_1=\frac{\Gamma(1/4)^8}{6 (2\pi)^{4}} \ ,
\end{align}
and $c$ is defined in (\ref{lifmi}).

At high temperature ($l/r_H\gg 1$), $r_c$ approaches $r_H$. Equation (\ref{lifee}), does not converge for $r_c=r_H$; we will rewrite equation (\ref{lifee}) in a way that allows us to take the limit $r_c\rightarrow r_H$ without encountering any divergence
\begin{align}
A=\frac{2L}{a}+ \frac{L}{r_c}\left[-\frac{(2\pi)^{3/2}}{\Gamma(1/4)^2}+\frac{\Gamma(-1/4)\Gamma(1/2)}{2\Gamma(1/4)}+\frac{l}{r_c}+ \sum_{n=1}^\infty \frac{1}{2n-1}\frac{\Gamma\left[\frac{1}{2}+n\right]\Gamma \left[\frac{n}{2}+\frac{3}{4}\right]}{ \Gamma[1+n]\Gamma \left[\frac{n}{2}+\frac{5}{4}\right]}\left(\frac{r_c}{r_H}\right)^{2n}\right] \ .
\end{align}
Now the entanglement entropy of the rectangular strip for the boundary theory at high temperature ($l\sqrt{T/R}\gg 1$) is obtained by taking the limit $r_c\rightarrow r_H$ in the last equation, yielding 
\begin{equation} \label{higheelif}
S_A=c \, \left[\frac{2L}{a} +\frac{2\pi L l T}{R}- \L_{\rm high}\frac{L\sqrt{T}}{\sqrt{R}}+ \ldots \right] \ ,
\end{equation}
where, $\L_{\rm high}=2.671$.

\renewcommand{\theequation}{C.\arabic{equation}}
\setcounter{equation}{0}  
\section*{Appendix C. Computations in the Hyperscaling-violating background}
\addcontentsline{toc}{section}{Appendix C. Computations in the Hyperscaling-violating background}

\noindent{\bf The case of $\theta \not = d-2$:}
\begin{align}
l=&r_c \sum_{n=0}^{n=\infty}p_n \left(\frac{r_c}{r_H}\right)^{n\gamma},\\
S_{A; \rm finite}=&\frac{2 c\, L^{d-2}}{r_c^{d-\theta-2}} \left[q_0+ \sum_{n=1}^{n=\infty}q_n \left(\frac{r_c}{r_H}\right)^{n\gamma}\right] \ ,
\end{align}
where, $p_n,q_n$ are constants that depend only on $d$ and $\theta$ and $c$ is defined in (\ref{hsmi}). At low temperature $r_c\ll r_H$ and we get
\begin{equation}
 S_{A; \rm finite}=\frac{c\, \C(\theta,d) L^{d-2}}{l^{d-\theta-2}}\left[1+h_1~ l^{\gamma}~ T^{\frac{\gamma}{z}}+...\right] \ , 
\end{equation}
where, $h_1$ is a numerical constant and 
\begin{align}
\C(\theta,d)=2 p_0^{d-\theta-2} q_0 \ .
\end{align}
At high temperature $r_c\sim r_H$ and our previous calculations suggests that in the limit $r_c\rightarrow r_H$ we can write
\begin{equation}
 S_{A; \rm finite}=\frac{2 c\, L^{d-2}}{r_H^{d-\theta-2}} \left[q_0-p_0+\frac{l}{r_H}+ \sum_{n=1}^{n=\infty}(q_n-p_n) \right]
\end{equation}
and the infinite sum now converges. Finally, we obtain
\begin{equation} \label{eehighhs}
 S_{A; \rm finite}= c  \,  L^{d-2}~ T^{\frac{d-\theta-1}{z}}\left[h_2 l  +h_3 ~ T^{-\frac{1}{z}}+...\right] \ , 
\end{equation}
where, $h_2, h_3$ are numerical constants.\\

\noindent{\bf The case of $\theta = d-2$:} At low temperature $r_c\ll r_H$ and we get
\begin{equation}
 S_{A; \rm finite}=2L^{d-2} c \left[ \ln(l)+k_1 l^{\gamma} ~T^{\gamma/z}+...\right] \ ,
\end{equation}
where $k_1\ge 0$ is a numerical constant. At high temperature $r_c\sim r_H$ and we obtain
\begin{equation}
 S_{A; \rm finite}=c \,  L^{d-2}\left[k_2- \frac{2}{z}\ln(T)+k_3 l T^{1/z}+...\right] \ ,
\end{equation}
where, $k_2$ and $k_3$ are again numerical constants.

\end{document}